\documentclass[preliminary,copyright,creativecommons]{eptcs}
\providecommand{\event}{LFMTP 2023} 

\usepackage{iftex}

\ifpdf
  \usepackage{underscore}         
  \usepackage[T1]{fontenc}        
\else
  \usepackage{breakurl}           
\fi

\usepackage{graphicx}

\usepackage{amssymb}
\usepackage{graphicx}
\usepackage{proof}
\usepackage{amsthm}
\usepackage{mathtools}
\usepackage{multicol}
\usepackage[shortlabels]{enumitem}
\usepackage{latexsym}
\usepackage{listings}
\usepackage{srcltx}
\usepackage{comment}
\usepackage{xcolor}
\usepackage{xspace}
\usepackage{array}
\usepackage{stmaryrd}
\usepackage{float}
\usepackage{hhline}
\usepackage{url}
\usepackage{turnstile}
\usepackage{hyperref}

\newcommand{\bel}{\textsc{Beluga }}
\newcommand{\belend}{\textsc{Beluga}}
\newcommand{\har}{\textsc{Harpoon }}
\newcommand{\harend}{\textsc{Harpoon}}

\newcommand{\unifR}[3] {#1 ; #2 \xRightarrow{R} #3}
\newcommand{\unifL}[3] {#1 ; #2 \gg #3 \xRightarrow{L} \ceil{\Psi \vdash P}}
\newcommand{\focus}[3] {#1 ; #2 > #3 \Rightarrow \ceil{\Psi \vdash P}}
\newcommand{\lfunif}[2][\Psi] {\Delta ; #1 \xRightarrow{u} #2}
\newcommand{\lffocPsi}[2] {\Delta ; \Psi > #1 \rightrightarrows #2}

\newcommand{\sequent}[3][\Delta] {#1 ; #2 \Longrightarrow #3}
\newcommand{\lfseq}[3][\Delta] {#1 ; #2 \Longrightarrow #3}

\newcommand{\lfnormtyping}[3][\Psi] {\Delta ;#1 \vdash #2 \Leftarrow #3}

\newcommand{\msub}{[\![ C/X ]\!]}

\newcommand{\posG}{\Gamma^{-}_{\Delta, \Gamma}}
\newcommand{\negD}{\Delta^{+}_{\Delta, \Gamma}}
\newcommand{\alp}{\ceil{\Psi \vdash P}}

\newtheoremstyle{custom}
                {\topsep}
                {\topsep}
                {\itshape}
                {}
                {\bfseries}
                {.}
                {\newline}
                {}

\theoremstyle{custom}
\newtheorem*{theorem*}{Theorem}
\newtheorem{theorem}{Theorem}
\newtheorem{lemma}{Lemma}

\usepackage[lighttt]{lmodern}

\newcommand{\nicettfamilysize}{\footnotesize}

\newcommand{\nicettfamily}{%
  \nicettfamilysize\ttfamily\linespread{1}\selectfont%
}%

\lstdefinelanguage{Beluga}{%
  morecomment = [l][\color{Gray}]{\%},
  morekeywords =[1]{
    LF, type, ctype, schema, rec, proof, as, case, by, of, unbox, inductive, %
    stratified, some, block, total, mlam, fn, let, in, box, %
    intros, solve, msplit, suffices, split, by, as, invert, case, unboxed, %
    impossible, undo, toshow, auto, %
    strengthen, %
  },%
  keywordstyle = [1]{\bfseries},%
  morekeywords =[2]{
    arr, app, abs, b, beta, stepapp, id, sstep, c, halts/m, val/c, val/abs, tp_c, tp_abs, tp_app, ev_c, ev_app, ev_abs%
  },%
  keywordstyle = [2]{\color{blue}},%
  morekeywords = [3]{
    tp, term, val, step, steps, halts, oft, eval, exp %
  },%
  keywordstyle = [3]{\color{magenta}},%
  morekeywords = [4]{
    I, Arr, halts_step, bwd_closed, pres%
  },%
  keywordstyle = [4]{\color{teal}},%
  morekeywords = [5]{
    Reduce%
  },%
  keywordstyle = [5]{\color{purple}},%
  alsoletter=/,%
  columns=flexible,%
  sensitive = true,%
  basicstyle=\nicettfamily,%
  mathescape=true,%
  escapechar={*},
  literate={%
    {→}{$\rightarrow$ }1%
    {⇒}{$\Rightarrow$ }1%
    {->}{$\rightarrow$ }1%
    {\\}{$\lambda$}1%
    {⊢}{$\vdash\;$}1%
    {|-}{$\vdash\;$}1%
    {sigma}{$\sigma$}1%
  }%
}%

\newcommand\ceil[1]{\lceil#1\rceil}
\newcommand{\cbox}[1]{\ceil {#1}}
\newcommand{\sid}{\ensuremath{\mathsf{wk}}}

\newcommand{\mlam}{\ensuremath{\lambda^{\square}}\xspace}
\newcommand{\tfn}{\ensuremath{\textsf{fn}}~}
\newcommand{\tbox}{\ensuremath{\textsf{box}}\xspace}
\newcommand{\tletbox}{\ensuremath{\textsf{let box}}}

\newcommand{\boxit}[1]{\tbox\ (#1)}
\newcommand{\letbox}[3]{\tletbox\ {#1} = #2\  \textsf{in}\ #3}
\newcommand{\mapp}[2]{#1~(#2)}

\newcommand{\Psihat}{\widehat{\Psi}}

\title{Semi-Automation of\\ Meta-Theoretic Proofs in Beluga}
\author{Johanna Schwartzentruber
\institute{McGill University\\ Montreal, Canada}
\email{johanna.schwartzentruber@mail.mcgill.ca}
\and
 Brigitte Pientka
\institute{McGill University\\ Montreal, Canada}
\email{ bpientka@cs.mcgill.ca}
}

\def\titlerunning{Automating Proof Search in Beluga}
\def\authorrunning{J. Schwartzentruber \& B. Pientka}
\begin{document}
\maketitle

\begin{abstract}
We present a sound and complete focusing calculus for the core of the logic behind the proof assistant \bel as well as an overview of its implementation as a tactic in \belend's interactive proof environment \harend. The focusing calculus is designed to construct uniform proofs over contextual LF and its meta-logic in \belend: a dependently-typed first-order logic with recursive definitions. The implemented tactic is intended to complete straightforward sub-cases in proofs allowing users to focus only on the interesting aspects of their proofs, leaving tedious simple cases to \belend's theorem prover. We demonstrate the effectiveness of our work by using the tactic to simplify proving weak-head normalization for the simply-typed lambda-calculus.
\end{abstract}

\section{Introduction}
To establish trust in a software system, we need to begin with establishing trust in
the programming language (PL) that is used in its implementation. 
To model formal systems like PLs, we first need to
specify them in a specification logic, and then prove statements about
the behaviour of the specified system within another logic, called the
reasoning logic. A key challenge in specifying languages is how to model
variable bindings, variable renaming, $\alpha$-conversion,
substitution, and reasoning under assumptions.
Higher-order abstract syntax (HOAS) offers a solution to these issues, reducing the amount of overhead infrastructure users must construct.
On the reasoning side, proofs about these language specifications for
even small languages
can still involve long and tedious proofs where many sub-cases are
trivial, and the only challenging aspect pertains to one or two
interesting parts. Our goal is to ease the burden on specifying and
proving properties by providing users with automated support based on proof theoretic foundations to
discharge trivial, straightforward cases automatically.

There are several HOAS-based systems that offer varying degrees of
automation. Abella \cite{abellaSystemdesc} allows users to develop
proofs interactively using a small set of tactics which offer little
automation. Hybrid \cite{first-hybrid, official-hybrid} adds HOAS
support to the general proof assistants Coq \cite{coq} and
Isabelle/HOL \cite{Isabelle} and consequently is able to harness
their automation and tooling powers, at the cost of added work to
specify systems \cite{comparison-HOAS, 700challenges-pt2}. Twelf
offers full automation; although users are not able to interact with
its prover which does not support backtracking. Most importantly,
Twelf and Abella do not produce proof terms and therefore offer no way
to verify their proofs independently.

We investigate proof automation within the HOAS-based proof assistant, \bel
\cite{bel-sys-desc}. 
In \belend, users formalize their systems within the logical framework LF
\cite{eLF} and subsequently prove properties about LF objects in
\belend’s reasoning logic: a dependently typed first-order logic.
\bel takes a functional approach; modelling inductive proofs about LF objects as recursive dependently-typed programs, following the Curry-Howard isomorphism. To ease proof developments, a tactic-based prover \har was recently added \cite{harpoon-interactive}. Users may now construct proofs interactively using a small set of high-level actions similar to those in Abella, and, upon completion of a proof, will be presented with a proof script which can be translated to a \bel program that may be independently type-checked. It is still the case however that much human interaction is required for \har proof developments.

In this paper we present a focusing calculus designed to build uniform proofs over the core of \belend's two-level logic, and provide an overview of its implementation as a \har tactic.
The tactic is meant to solve simple lemmas and simple cases of proofs, allowing users to
concentrate on the interesting aspects of a proof.
We have proven that the focusing calculus, presented in Section \ref{focusingChapter}, is sound and complete with respect to the cut-free sequent calculus for \belend, presented in Section \ref{seqCalc}.
 We have used our tactic on a number of interesting case studies in
 PL theory, which we summarize in Section \ref{eval}, including type preservation and value soundness for MiniML
 (without fix points), weak-head normalization for the simply-typed
 lambda-calculus, and the Church-Rosser theorem for the untyped
 lambda-calculus. We have seen that they allow for automatic
 completion of many of the simpler lemmas and subcases of these
 theorems. We believe automating proof search over the core of \bel
 results in simpler proof developments, making it more appealing to users looking to verify formal systems.

\section{Introduction to \belend}

In this section, we present an overview of the \bel system. We begin with an informal, followed by a formal, description of its logic, concluding with a presentation of the focusing calculus that we implement.

\subsection{Encoding (meta-)theories}
\label{theory}

We introduce \bel informally by demonstrating how theories and their meta-theories are encoded within the system. Throughout this paper we will focus on the proof development of a key lemma that is used to prove weak-head normalization for the simply-typed lambda-calculus using logical relations. The lemma states that reducibility is closed under expansion, i.e. if term \lstinline[columns=fixed]{M} steps to term \lstinline[columns=fixed]{N} and \lstinline[columns=fixed]{N} reduces to type \lstinline[columns=fixed]{A}, then \lstinline[columns=fixed]{M} does as well. A detailed description of the full mechanization may be found at \cite{full-mech}. 

We begin by encoding the theory of the simply-typed lambda-calculus
within \belend's specification logic, the logical framework LF
\cite{eLF} making use of HOAS. We choose an intrinsically-typed representation to simplify our mechanization.

{
  \centering
\begin{minipage}{0.4\linewidth}
\begin{lstlisting}[xleftmargin=.15\textwidth]
LF tp : type =
| b :  tp
| arr : tp -> tp -> tp ;
\end{lstlisting}
  \end{minipage}
  \begin{minipage}{0.6\linewidth}
\begin{lstlisting}
LF term : tp -> type =
| app : term (arr A B) -> term A -> term B
| abs : tp -> (term A -> term B) -> term (arr A B)
| c : term b ;
\end{lstlisting}
 \end{minipage}
}

Next, we define the operational semantics of our lambda terms. For
simplicity we do not reduce under abstractions. We observe that object-level substitution is modelled by LF application, as in the type of \lstinline[columns=fixed]{beta}.

{
  \centering
\begin{minipage}{0.54\linewidth}
\begin{lstlisting}[xleftmargin=.05\textwidth]
LF step : term A → term A → type =
| beta : step (app (abs A M) N) (M N)
| stepapp : step M M' 
             -> step (app M N) (app M' N) ;
\end{lstlisting}
  \end{minipage}
  \begin{minipage}{0.5\linewidth}
      \begin{lstlisting}
LF steps : term A -> term A -> type =
| id   : steps M  M
| sstep : step M M' 
           -> steps M' M'' -> steps M M'' ;
\end{lstlisting}
  \end{minipage}
}

To complete the theory's encoding, we also define what it means for a term to halt, i.e. it steps to a value. 

{
  \centering
\begin{minipage}{0.45\linewidth}
\begin{lstlisting}[xleftmargin=.15\textwidth]
LF val : term A -> type =
| val/c : val c
| val/abs : val (abs A M) ;
\end{lstlisting}
  \end{minipage}
  \begin{minipage}{0.5\linewidth}
\begin{lstlisting}
LF halts : term A -> type =
| halts/m : steps M M' -> val M' -> halts M ;
\end{lstlisting}
 \end{minipage}
}


To reason about LF objects, we embed them within the computation logic
using a modal box (necessity) operator  \cite{POPL08}. Users also have
the ability to reason about ``open'' LF objects in \belend. To do
this, we pair each LF object together with the LF context in which it
is meaningful \cite{POPL08, PPDP08}. This concept is internalized as a
contextual type $\cbox{\Psi \vdash P}$ \cite{CMTT}. This contextual
type describes a computation-level expression $\boxit{\Psihat \vdash R}$ where
$R$ is an LF object of type $P$ in the context $\Psi$. In other words,
$\Psihat$ describes the free variables in $R$ and corresponds to
the erased typing context $\Psi$. 

We present a trivial result about our specified theory to demonstrate how meta-theorems are expressed in \belend.

{
\centering
\begin{minipage}{1\linewidth}
\begin{lstlisting} [xleftmargin=.09\textwidth]
rec halts_step : [ |- step M M'] -> [ |- halts M'] -> [ |- halts M] =
fn s, h =>
   let [ |- halts/m MS V] = h in let [ |- S] = s in [ |- halts/m (sstep S MS) V] ;
\end{lstlisting}
\end{minipage}
}

We leave the contextual variables \lstinline[columns=fixed]{M}, \lstinline[columns=fixed]{M'}, and \lstinline[columns=fixed]{M''} implicitly universally quantified as \bel is able to reconstruct their type. We use \belend's simple function space to formalize our implication statement. Recall proofs are programs in \belend, therefore proof development proceeds in a functional manner.

We begin by giving the theorem name and statement, prefixed with the keyword \lstinline[columns=fixed]{rec}. The proof starts by peeling off the implication antecedents (\lstinline[columns=fixed]{fn s, h =>}). Working backwards, we know we must use \lstinline[columns=fixed]{halts/m} to construct our desired term as it is currently the only constructor for terms of type \lstinline[columns=fixed]{halts}, therefore we must solve its subgoals, namely that there is a value that \lstinline[columns=fixed]{M} steps to. We first invert \lstinline[columns=fixed]{h} as it has only one possible constructor. This reveals that it is actually the contextual object \lstinline[columns=fixed]{[ |- halts/m MS V]} where \lstinline[columns=fixed]{MS} and \lstinline[columns=fixed]{V} are LF terms of type \lstinline[columns=fixed]{steps M' N} and \lstinline[columns=fixed]{val N} (for some implicit meta-variable \lstinline[columns=fixed]{N}) respectively. It may appear we have every piece of the puzzle required to solve our goal: we have a value that our term \lstinline[columns=fixed]{M} steps too. However once we transition to the LF level to build our LF proof term, we do not have access to our computation-level context, in which \lstinline[columns=fixed]{s} resides. Therefore, we must first \textit{unbox} said assumption. 
 


The free variables appearing in all the specifications above are treated as implicitly quantified. \bel infers their types during type reconstruction. As such, users do not supply arguments for such parameters.


%
%

\subsection{Theoretical foundation}

We give a formal presentation of the core of \belend's two-level logic
beginning with its grammar followed by its two-level proof system described using two cut-free sequent calculi. For a full description of the logic, readers may refer to past works \cite{POPL12, POPL08}.

\subsubsection{Grammar}

\belend's two-level logic is based on dependent contextual modal type
theory (CMTT) \cite{CMTT}. At the core of the specification logic is the logical
framework LF \cite{eLF} which supports encodings using HOAS. 
We give here a definition that characterizes only
canonical (normal) objects as these are the only ones that are
meaningful in our setting.

We separate terms into two categories, neutral and normal. We characterize neutral terms to be those that do not cause beta-redexes when they are applied in function application. 
Terms are classified by types, which are either type constants \textbf{a} that may be indexed by terms $M_1,\dots,M_n$, or dependent types. 
\vspace{-0.04 cm}
\[
\begin{small}
\begin{array}[h]{llrlllrl}
\text{Atomic Types } & P,Q & ::= & \textbf{a} \overrightarrow{M} & \text{Substitutions } & \sigma &::= & \cdot \mid \sid_\psi \mid \sigma, M
\\
\text{Types } & A,B & ::= & P \mid \Pi x{:}A.\: B & \text{Contexts } & \Psi, \Phi &::= & \cdot \mid \Psi, x:A \mid \psi
\\
\text{Neutral Terms } & R &::= & x \mid \textbf{c} \mid R\ N \mid u[\sigma] & \text{Contextual Variables } & X &::= & u[\sigma] \mid \psi
\\
\text{Normal Terms } & M,N &::= & R \mid \lambda x.\: M
\end{array}
\end{small}
\]

To support pattern matching on LF objects, we further extend LF with
two kinds of contextual variables: meta-variables, written as
$u[\sigma]$ and context variables, written as $\psi$. 
Context variables allow for abstraction over contexts which is required for recursion over HOAS specifications. 
 Meta-variables allow us to describe ``holes'' in an LF object. They
 describe  possibly open objects that are paired with a postponed
 simultaneous substitution $\sigma$ (by convention written to the
 right of a term) that is applied as soon as we know what $u$ stands for. 

Simultaneous substitutions $\sigma$ provide a mapping from one context of variables $\Phi$ to another $\Psi$. We do not always make the domain of the substitution explicit, but one can think of the i-th element of $\sigma$ corresponding to the i-th declaration in $\Phi$. We assume all substitutions are hereditary substitutions \cite{hered-subst}.

Variables in a contextual LF expression may be bound by one of two contexts. There is the LF context $\Psi$ that holds typings for ordinary variables, and there is the meta-context $\Delta$ (introduced below) which holds typings for contextual variables, uniformly denoted by $X$. Contextual variables include meta-variables and context variables $\psi$. 

In order to uniformly abstract over meta-objects in the computation logic, we lift contextual LF objects to meta-types $U$ and meta-terms $C$. 

\vspace{-0.5 cm}
\[
\hspace{-0.1cm}
\begin{small}
\begin{array}[h]{llrlllrl}
\text{Meta Terms } & C &::= & (\hat{\Psi} \vdash R) \mid \Psi & \text{Context Schemas } & G &::= &  \exists \overrightarrow{(x{:}A_o)}.\ A \mid G + \exists \overrightarrow{(x{:}A_o)}.\ A
\\
\text{Meta Types } & U &::= & (\Psi \vdash P) \mid G & \text{Meta Contexts } & \Delta &::= & \cdot \mid \Delta, X : U  \\
\text{Meta Substitutions } & \theta &::= & \cdot \mid \theta, C/X 
\end{array}
\end{small}
\]

The core of our meta language's terms include contextual terms as well
as LF contexts. The meta-type $(\Psi \vdash P)$ denotes the type of a
meta-variable $u$ and stands for a contextual term. For simplicity, we
restrict context schemas $G$ to be constructed from schema elements
$\exists \overrightarrow{(x {:} A_o)}.A$ using $+$, where $A$ is an LF
type. 

We write the single meta-substitution as $[\![ C / X]\!]$. In most
cases $X$ stands for a meta-variable $u$ and $C$ stands for a
contextual object $(\hat{\Psi} \vdash R)$. In this case the
substitution gets pushed through $\lambda$-expressions until we reach
a meta-variable $u[\sigma]$. We then apply the meta-substitution to its
associated substitution to obtain $\sigma'$ before eagerly applying $\sigma'$
to $R$. The full definition of meta-substitutions may be found at \cite{POPL12, POPL08}.

On top of contextual LF we have the computational layer, which is used
to describe programs that operate on data. The computation types
include atomic box-types $\cbox{\Psi \vdash P}$, computation level function
abstraction, as well as abstraction over contextual objects. 

\vspace{-0.04 cm}
\[
\begin{small}
\begin{array}{llrl}
\text{Types } & \tau & ::= & \cbox{\Psi \vdash P} \mid \tau_1 \rightarrow \tau_2 \mid \Pi^{\Box} X{:}U. \: \tau 
\\
\text{Expressions} & E & 
::= & y 
\mid \boxit{\hat{\Psi} \vdash R}
\mid \letbox X {E_1} E_2 \\
 & & & \mid \tfn  y. E
\mid E_1\ E_2
\mid \mlam X. E 
\mid \mapp{E}{C}
\\
\text{Computation-level Contexts } & \Gamma & ::= & \cdot \mid \Gamma, y:\tau
\end{array}
\end{small}
\]

Ordinary functions are
created using $\tfn y. E$, while we write 
$\mlam X. E$ for 
dependent functions that abstract over meta-objects.
We overload the application operation. We write $E_1~E_2$ for applying the
expression $E_1$ of function type $\tau_1 \rightarrow \tau_2$ to an expression $E_2$. We also write $\mapp{E}{C}$ for applying the expression
$E$ of type $\Pi^{\Box} X{:}U. \: \tau$ to a contextual object $C$.



\subsubsection{Sequent calculi}
\label{seqCalc}

We now present the sequent calculus for contextual LF, which is based
on the sequent calculus for intuitionistic contextual modal logic
presented in \cite{CMTT}. 
Following the logic programming
interpretation of LF, a proof of a proposition encoded as an LF type
is an LF term which inhabits said type. 
We exploit the fact that if
$B$ does not depend on $x$ in $\Pi x{:}A.B$, we interpret it as an
implication $A \rightarrow B$. If $x$ does occur in $B$, then we treat
$\Pi x{:}A.B$ as universal quantification, written as $\Pi
\hat{x}{:}A.B$.

 All sequents have access to the global signature $\Sigma$ which we keep implicit. The sequent $\lfseq{\Psi}{M:A}$ states that $M$ is a proof of the proposition $A$, or $M$ has type $A$, using assumptions from the meta-context
 $\Delta$ and the LF context $\Psi$. As a consequence, we keep in the
 LF context $\Psi$ not only proof-relevant assumptions that arise from
 implications (in particular $\rightarrow R$), but we also add
 parameter assumptions that come from $\Pi R$. We distinguish between
 these two assumptions by writing $x{:}A$ for the former, and
 $\hat{x}:A$ for the latter. 
 This is emphasized in the init$^\Psi$ rule, where only proof-relevant 
assumptions are used to finish proofs. 

In addition to the sequent $\lfseq{\Psi}{M: A}$, we also have
$\lfseq{\Psi}{\sigma:\Phi}$ which states that the substitution $\sigma$ witnesses a proof of
the propositions in $\Phi$ using assumptions from $\Delta$ and
$\Psi$. Intuitively, $\Phi$ is an LF context containing assumptions
$x_i{:}A_i$ and $\hat{x}_j{:}A_j$. Note that we need to construct a
proof for the former, but for the latter we are able to determine the
witness via unification in practice.  



\begin{figure}[h]
  \begin{small}
\[
\begin{array}[h]{c}
\multicolumn{1}{l}{\fbox{$\lfseq{\Psi}{M:A}$}~\mbox{Object $M$ is a proof
  term for the proof of $A$ in the sequent calculus}}\\[1em]
\infer[\text{init$^{\Sigma}$}]{\lfseq{\Psi}{\textbf{c}: A}}{\textbf{c}:A \in \Sigma}
\qquad
\infer[\text{init$^{\Psi}$}]{\lfseq{\Psi, x: A}{x: A}}{}
\qquad
\infer[\Pi R]{\lfseq{\Psi}{\lambda x. M : \Pi \hat{x}{:}A. \:
  B}}{\lfseq{\Psi, \hat{x}:A}{M: B}}
\\[1em]
\infer[\Pi L] {\lfseq{\Psi, x_1: \Pi \hat{x}{:}A. \: B}{[x_1 M/ x_2]N: A'}}{\lfnormtyping[\Psi, x_1: \Pi \hat{x}{:}A.\ B]{M}{A} & \lfseq{\Psi, x_1: \Pi \hat{x}{:}A. \: B, x_2: [M/\hat{x}] B}{N: A'}}
\\[1em]
\infer[\rightarrow R]{\lfseq{\Psi}{\lambda x. M : A \rightarrow B}}{\lfseq{\Psi, x:A}{M: B}}\quad
\infer[\rightarrow L] {\lfseq{\Psi, x_1: A \rightarrow B}{[x_1 M/x_2] N: A'}}{\lfseq{\Psi, x_1: A \rightarrow B}{M : A} & \lfseq{\Psi, x_1: A \rightarrow B, x_2: B}{N: A'}}
\\[1em]
\infer[\text{reflect}]{\lfseq[\Delta, u: (\Phi \vdash
  P)]{\Psi}{[u[\sigma]/x]M: A}}{\lfseq[\Delta, u: (\Phi \vdash
  P)]{\Psi}{\sigma: \Phi} & \lfseq[\Delta, u: (\Phi \vdash P)]{\Psi,
                            x: [\sigma]P}{M: A}}
\\[1em]
\multicolumn{1}{l}{\fbox{$\lfseq{\Psi}{\sigma:\Phi}$}~\mbox{Object
  $\sigma$ is a substitution that witnesses the proof of $\Phi$ in the sequent calculus}}\\[1em]
\infer[\text{sub}_{\text{empty}}]{\lfseq{\Psi}{ \cdot: \cdot}}{}
\qquad
\infer[\text{sub}_{\sid}] {\lfseq{\psi, \Psi}{\sid_{\psi}: \psi}} {}
\\[1em]

\infer[\text{sub}_p] {\lfseq{\Psi}{ (\sigma, N): (\Phi, x{:}B)}} {\lfseq{\Psi}{\sigma: \Phi} & \lfseq{\Psi}{N: [\sigma]B}}
 \qquad
\infer[\text{sub}_u] {\lfseq{\Psi}{ (\sigma, M): (\Phi, \hat{x}{:}B)}}
  {\lfseq{\Psi}{\sigma: \Phi} & 
  \lfnormtyping[\Psi]{M}{[\sigma]B}}
\end{array}
\]    
  \end{small}
\caption{Sequent calculus for contextual LF.}
\label{fig:seqCalcLF}
\end{figure}

The rules for the sequent calculus for contextual LF are presented in Figure \ref{fig:seqCalcLF}, on the next page. The right rules introduce variable declarations into the local context $\Psi$. However, those introduced via the $\Pi R$ rule are simply parameters that are not used during proof search, unlike those introduced via $\rightarrow R$. To use a universally quantified assumption (i.e. a dependent function type) as in $\Pi L$ we require that $M$ checks against type $A$, written $\lfnormtyping{M}{A}$, in the appropriate contexts. In practice, we do not search for the term $M$ but introduce meta-variables for such universally quantified variables which are later instantiated via unification. In contrast, using an assumption of ordinary function type, as in $\rightarrow L$, involves searching for a proof term of type $A$. Note that in the left rules we replace a neutral term by a neutral term, thus still constructing normal proofs.

In the reflect rule we may use the contextual assumption $(\Phi \vdash
P)$ to deduce $P$ in the context $\Psi$ if we can verify $\Phi$. In
order to verify $\Phi$ we need to find a substitution which maps all
the variables in $\Phi$ to terms that make sense in $\Psi$. There are
several ways to construct such a substitution, depending on the shape
of $\Phi$. If it is empty, we simply use an empty substitution (as $P$
is closed). If it is a context variable $\psi$ and we simply want to
use $(\psi \vdash P)$ in a weaker context, we apply the weakening
substitution. Otherwise, $\Phi$ contains two different kinds of
variable declarations: a declaration $x{:}B$  requires proof search in
order to find a term $N$ in $\Psi$ of type $[\sigma]B$ to replace $x$
in $P$;
a declaration $\hat{x}{:}A$ stands for a universally quantified
variable and does not require proof search. In practice, we can
determine it via unification. 
This different treatment of assumptions reflects their different roles.

\begin{figure}[h]
  \begin{small}
\[
\begin{array}[h]{c}
\multicolumn{1}{l}{\fbox{$\sequent{\Gamma}{E: \tau}$}~\mbox{Object $E$ is a proof
  term for the proof of $\tau$ in the sequent calculus}}\\[1em]
\infer[\Pi^{\Box} R]{\sequent{\Gamma}{\mlam X. E: \Pi^{\Box} X{:}U. \: \tau}}{\sequent[\Delta, X:U]{\Gamma}{E: \tau}}
\quad
\infer[\Pi^{\Box} L] {\sequent{\Gamma, y_1: \Pi^{\Box} X{:}U. \: \tau'}{[\mapp{y_1}{C}/y_2] E: \tau}}{\Delta \Vdash C \Leftarrow U & \sequent{\Gamma, y_1: \Pi^{\Box} X{:}U. \: \tau', y_2: [\![ C/X ]\!] \tau'}{E: \tau}}
\\[1em]
\infer[\rightarrow L] {\sequent{\Gamma, y_1: \tau_1 \rightarrow \tau_2}{[y_1 \ E'/y_2] E: \tau}}{\sequent{\Gamma, y_1: \tau_1 \rightarrow \tau_2}{E' : \tau_1} & \sequent{\Gamma, y_1: \tau_1 \rightarrow \tau_2, y_2: \tau_2}{E: \tau}} 
\\[1em]
\infer[\rightarrow R]{\sequent{\Gamma}{\tfn y.\ E : \tau_1 \rightarrow \tau_2}}{\sequent{\Gamma, y: \tau_1}{E: \tau_2}}
\qquad
\infer[\text{init}]{\sequent{\Gamma, y: \tau}{y: \tau}}{}
\quad
\infer[\Box R]{\Delta; \Gamma \Longrightarrow \boxit{\hat{\Psi} \vdash R} : \cbox{\Psi \vdash P}}{\Delta; \Psi \Longrightarrow R : P}
\\[1em]
\infer[\Box L] {\sequent{\Gamma, y{:} \cbox{\Psi \vdash P}}{\letbox{X}{y} E: \tau}}{\sequent[\Delta, X{:} (\Psi \vdash P)]{\Gamma, y{:}\cbox{\Psi \vdash P}}{E: \tau}}
\end{array}
\]    
  \end{small}
\caption{Sequent calculus for the computation logic.}
\label{fig:seqCalcC}
\end{figure}

We turn our attention to proof search over computations (see Figure \ref{fig:seqCalcC}). Our inference rules are mostly standard for a first-order logic. The $\Box R$ rule is the transition rule between contextual LF and computation-level proofs. In $\Box L$ we \textit{unbox} a boxed assumption, adding it to $\Delta$. Using computation assumptions as in $\Pi^{\Box} L$ and $\rightarrow L$ is similar to how contextual LF assumptions are used. To use a universally quantified assumption, as in $\Pi^{\Box} L$, we require that $C$ checks against $U$. Again, this term $C$ is not explicitly constructed but found instead through unification. We note that meta-objects may only depend on the meta-context, hence we only require that $C$ check against type $U$ in $\Delta$.

We show in \cite{my-thesis} that cut and contextual cut are admissible in the computation logic. These results are shown for contextual LF in \cite{CMTT}. Using the admissibility of cut results, we can deduce invertibility of some of the rules in our calculi. Interestingly, the box constructor $\Box$ has an invertible \textit{left} rule, which will have implications in the focusing calculus. We omit the proof terms for readability here.

\begin{lemma}[Invertibility in the sequent calculi]\leavevmode
    a)\ ($\Pi R$) If $\Delta; \Psi \Longrightarrow \Pi \hat{x}{:}A. B$ then $\Delta; \Psi, \hat{x}:A \Longrightarrow B$
    \\
    b)\ ($\rightarrow R$) If $\Delta; \Psi \Longrightarrow A \rightarrow B$ then $\Delta; \Psi, x:A \Longrightarrow B$
    \\
    c)\ ($\Pi^{\Box} R$) If $\Delta; \Gamma \Longrightarrow \Pi^{\Box} X{:}U. \tau$ then $\Delta, X:U; \Gamma \Longrightarrow \tau$
    \\
    d)\ ($\rightarrow R$) If $\Delta; \Gamma \Longrightarrow \tau_1 \rightarrow \tau_2$ then $\Delta; \Gamma, y: \tau_1 \Longrightarrow \tau_2$
    \\
    e)\ ($\Box L$) If $\Delta; \Gamma, y: \cbox{\Psi \vdash P} \Longrightarrow \tau$ then $\Delta, X:(\Psi \vdash P); \Gamma, y: \cbox{\Psi \vdash P} \Longrightarrow \tau$
\end{lemma}

\subsection{Focused proof system}
\label{focusingChapter}

As a foundation for automating proof search in \bel we develop a
focused sequent calculus over \belend's two-level logic. For the
description of this focused proof system, we omit proof terms when permitted to ease
readability. This logic formalizes the proof search procedure that we implement. The loop is fully automatic and therefore requires that non-determinism be handled with ease. The sequent calculus previously presented does not suffice as the rules do not provide any inherent direction for proof search. The rules of the following calculi guide better proof development. The calculus builds uniform proofs \cite{uniform} by applying all invertible rules first. We then handle non-invertible rules systematically through focusing \cite{Andreoli92}.

The focusing calculus for contextual LF consists of two main phases: a uniform and focusing phase, and is mostly straightforward (see Figure \ref{focLF}). The uniform proof phase consists of applying the invertible rules until we reach an atomic goal ($P$). During this phase, parameters and assumptions are collected and placed in the LF context. We then try to find a solution by focusing on assumptions from the different contexts. In particular, we iterate through assumptions (in the meta and LF contexts), decomposing each one into the atoms it defines without utilizing any other assumption.

\begin{figure}[h!]
\begin{small}
\[
\begin{array}[h]{c}
\multicolumn{1}{l}{\fbox{$\lfunif{A}$}~\mbox{There is a uniform proof of $A$ in the focusing calculus}}\\[1em]
\infer[\Pi R] {\lfunif{\Pi \hat{x}{:}A.\ B}} {\lfunif[\Psi, \hat{x}{:}A]{B}}
\qquad
\infer[\rightarrow R] {\lfunif{A \rightarrow B}} {\lfunif[\Psi, x:A]{B}}
\\[1em]
\infer[\text{transition$^{\Delta}$}] {\lfunif{P}} {\Delta(X) = (\Phi \vdash Q) & \lfunif{\sigma: \Phi}  & [\sigma]Q = P}
\qquad
\infer[\text{transition$^{\Psi}$}]{\lfunif{P}}{\Psi(x) = A & \lffocPsi{x: A}{P}}
\\[1em]
\multicolumn{1}{l}{\fbox{$\lfunif{\sigma: \Phi}$}~\mbox{Object $\sigma$ is a substitution
  that witnesses a uniform proof of $\Phi$ in the focusing calculus}}
\\[1em]
\infer[\text{empty}]{\lfunif{\cdot: \cdot}}{}
\qquad
\infer[\text{sub}_{\sid}]{\lfunif[\psi, \Psi]{\sid_{\psi}: \psi}}{}
\\[1em]
\infer[\text{sub}_p]{\lfunif{(\sigma, N): (\Phi, x{:}B)}}{\lfunif{\sigma: \Phi} & \lfunif{N: [\sigma]B}}
\qquad
\infer[\text{sub}_u]{\lfunif{(\sigma, M): (\Phi, \hat{x}{:}B)}}{\lfunif{\sigma: \Phi} & \lfnormtyping[\Psi]{M}{[\sigma]B}}
\\[1em]
\multicolumn{1}{l}{\fbox{$\lffocPsi{x: A}{P}$}~\mbox{There is a focused proof of $P$ with focus on $x$ in the focusing calculus}}
\\[1em]
\infer[\text{init$^{\Psi}$}]{\Delta;\Psi > x: P \rightrightarrows P}{}
\qquad
\infer[\rightarrow L]{\lffocPsi{x: A \rightarrow B}{P}}{\lfunif{A} & \Delta;\Psi > x': B \rightrightarrows P}
\\[1em]
\infer[\Pi L]{\lffocPsi{x: \Pi \hat{x}{:}A.\ B}{P}}{\Delta; \Psi \vdash M \Leftarrow A & \Delta;\Psi > x': [M/\hat{x}] B \rightrightarrows P} 
\end{array}
\]
\end{small}
\caption{Focusing calculus for contextual LF.}
\label{focLF}
\end{figure}

In the transition$^{\Delta}$ rule, using an assumption $(\Phi \vdash Q)$ from $\Delta$ to complete a proof requires a simultaneous substitution ($\sigma$) to be constructed so that $Q$ makes sense in the current LF context $\Psi$. We find such a substitution through proof search. When focusing on assumptions from $\Psi$ of function type, we search for a proof of $A$ if our assumption is of non-dependent function type. Otherwise the assumption is of dependent function type, in which case $M$ is found via unification.

Proof search over the reasoning layer proceeds similarly to LF (see Figure \ref{focComp}). We first perform all invertible rules, then we must make a choice on what to focus on. Unlike in LF proof search, proof search over computations requires two separate phases of inversions since the box connective has an invertible left rule. Further, in addition to focusing on the left, we also focus on the right, which corresponds to proof search in LF.

We begin with a uniform right phase which ends with an atomic goal formula, $\cbox{\Psi \vdash P}$. From there we transition to the uniform left phase where we unbox all box-type assumptions in $\Gamma$, moving them to $\Delta$. This is done because when we shift levels to LF proof search we only bring with us assumptions that are true across all levels (those in $\Delta$) and computation assumptions do not make sense on the LF level. The sequent for the uniform left phase is novel. We use the symbol $\gg$ as a way to distinguish assumptions that may be of box-type (to the right of $\gg$) from ones that are not (to the left of $\gg$). Recall that the order of assumptions in $\Gamma$ does not matter, therefore it is acceptable that the order reverses each time we complete a uniform left phase.

\begin{figure}[h]
\begin{small}
\[
\begin{array}[h]{c}
\multicolumn{1}{l}{\fbox{$\unifR{\Delta}{\Gamma}{\tau}$}~\mbox{There is a uniform right proof of $\tau$ in the focusing calculus}}\\[1em]
\infer[\rightarrow R]{\Delta;\Gamma \xRightarrow{\text{R}} \tau_1 \rightarrow \tau_2}{\Delta; \Gamma, y: \tau_1 \xRightarrow{\text{R}} \tau_2}
\qquad
\infer[\Pi^{\Box} R] {\Delta; \Gamma \xRightarrow{\text{R}} \Pi^{\Box} X{:}U. \: \tau} {\Delta, X:U; \Gamma \xRightarrow{\text{R}} \tau}
\qquad
\infer[\text{left to right}]{\Delta;\Gamma \xRightarrow{\text{R}} \alp}{\Delta; \cdot \gg \Gamma \xRightarrow{\text{L}} \alp }
\\[1em]
\multicolumn{1}{l}{\fbox{$\unifL{\Delta}{\Gamma}{\Gamma'}$}~\mbox{There is a uniform left proof of $\alp$ in the focusing calculus}}\\[1em]
\infer[\Box L] {\Delta; \Gamma \gg \Gamma ', y: \cbox{\Phi \vdash Q} \xRightarrow{\text{L}} \alp} {\Delta, X: (\Phi \vdash Q); \Gamma \gg \Gamma ' \xRightarrow{\text{L}} \alp}
\qquad
\infer[\text{shift}] {\Delta; \Gamma \gg \Gamma ', y: \tau \xRightarrow{\text{L}} \alp} {\tau \neq \cbox{\Phi \vdash Q} & \Delta; \Gamma, y: \tau \gg \Gamma ' \xRightarrow{\text{L}} \alp}
\\[1em]
\infer[\text{level}]{\Delta;\Gamma \gg \cdot \xRightarrow{L} \alp}{\lfunif{P}}
\qquad
\infer[\text{focus to uniform }] {\Delta; \Gamma \gg \cdot \xRightarrow{L} \alp} {\Gamma(y) = \tau & \Delta; \Gamma > y:\tau \Rightarrow \alp}
\\[1em]
\multicolumn{1}{l}{\fbox{$\Delta; \Gamma > y: \tau \Rightarrow \alp$}~\mbox{There is a focused proof of $\alp$ with focus on $y$ in the focusing calculus}}\\[1em]
\infer[\Pi^{\Box} L]{\Delta;\Gamma > y: \Pi^{\Box} X{:}U. \: \tau \Rightarrow \alp}{\Delta \Vdash C \Leftarrow U & \Delta;\Gamma > y': [\![C/X]\!] \tau \Rightarrow \alp}
\\[1em]
\infer[\rightarrow L]{\Delta;\Gamma > y: \tau_1 \rightarrow \tau_2 \Rightarrow \alp}{\Delta; \Gamma \xRightarrow{R} \tau_1 & \Delta;\Gamma > y': \tau_2 \Rightarrow \alp}
\qquad
\infer[\text{blur}]{\Delta;\Gamma > y':\cbox{\Phi \vdash Q} \Rightarrow \alp}{\Delta;\cdot \gg \Gamma, y': \cbox{\Phi \vdash Q} \xRightarrow{L} \alp}
\end{array}
\]
\end{small}
\caption{Focusing calculus for the computation logic.}
\label{focComp}
\end{figure}

Similarly to focusing in LF proof search, if we focus on a universally quantified assumption then we find $C$ through unification. Focusing on the right, i.e. LF proof search, can only be applied if the goal is of box-type. Focusing on the left is standard and commences once we have decomposed the focused formula into its atoms as in the blur rule. At this point, we add the atomic formula to $\Gamma$ and restart the process from the uniform left stage (to unbox the atomic formula if necessary). In practice, we implement backtracking when focusing. If for example, we cannot find a proof while focusing on the right, we backtrack and try focusing on the left. 
In practice we also support recursive types, which we treat as atomic computation-level types. These goal types may only be solved by focusing on the left.

We show in \cite{my-thesis} that the focusing calculi in Figures \ref{focLF} and \ref{focComp} are sound and complete with respect to the sequent calculi presented in Section \ref{seqCalc}. The completeness result in particular is interesting as it requires an intermediate result stating that it does not matter in which context box-type assumptions appear. That is, if there is a proof in our sequent calculus of $\cbox{\Psi \vdash P}$ (possibly) using some assumption $y: \cbox{\Phi \vdash Q}$ in $\Gamma$ then there is also a proof of $\cbox{\Psi \vdash P}$ where $y$ is omitted but under the added assumption $X: (\Phi \vdash Q)$ in $\Delta$. Given a $\Delta$ and $\Gamma$, $\posG$ denotes $\Gamma$ without any assumptions of box-type, and $\negD$ denotes $\Delta$ extended with the (unboxed) boxed assumptions from $\Gamma$.

\begin{theorem} [Soundness]
a) If $\lfunif{A}$ then $\sequent{\Psi}{A}$ \\
b)  If $\lfunif{\sigma: \Phi}$ then $\sequent{\Psi}{\sigma: \Phi}$ \\
c) If $\lffocPsi{x: A}{P}$ then $\sequent{\Psi,x: A}{P}$ \\
d) If $\Delta > X: U ; \Psi \rightrightarrows P$ then $\sequent[\Delta, X: U]{\Psi}{P}$ \\
e) If $\unifR{\Delta}{\Gamma}{\tau}$ then $\sequent{\Gamma}{\tau}$ \\
f) If $\unifL{\Delta}{\Gamma}{\Gamma'}$ then $\sequent{\Gamma,\Gamma'}{\alp}$ \\
g) If $\focus{\Delta}{\Gamma}{y: \tau}$ then $\sequent{\Gamma,y: \tau}{\alp}$
\end{theorem}

\begin{theorem}[Completeness]
a) If $\sequent{\Psi}{A}$ then $\lfunif{A}$ \\
b) If $\sequent{\Psi}{\sigma: \Phi}$ then $\lfunif{\sigma: \Phi}$ \\
c) If $\sequent{\Psi}{P}$ then $\lffocPsi{x: A}{P}$ for some $A \in \Psi$ or $\Delta > X: U; \Psi \rightrightarrows P$ for some $U \in \Delta$ \\
d) If $\sequent{\Gamma}{\tau}$ then $\unifR{\negD}{\posG}{\tau}$ \\
e) If $\sequent{\Gamma}{\alp}$ then $\unifL{\negD}{\posG}{\cdot}$ \\
f) If $\sequent{\Gamma}{\alp}$ then $\negD; \Psi \xRightarrow{u} {P}$ or $\negD; \posG > y: \tau \Rightarrow \alp$ for some $y: \tau \in \posG$
\end{theorem}

The soundness proofs proceed by mutual structural induction on the given derivations and are straightforward. Proving completeness however is more involved and depends on several minor lemmas, consisting mostly of postponement results and a variation of the reflect rule for computation-level focusing, all proven with straightforward induction. The completeness proof proceeds again by mutual structural induction on the given derivations. In part $a)$ the base case (i.e. the derivation is init$^{\Psi}$) is simplified by showing that the init$^{\Psi}$ rule is admissible under the addition of the rule $\lfseq{\Psi, x:P}{P}$. The remaining cases are then straightforward. Part $b)$ is trivial. Part $c)$ must be done by case analysis on the induction hypothesis but is otherwise straightforward. The base case in  part $d)$ is also made simpler by showing that the init$^{\Gamma}$ rule is admissible under the added assumption $\sequent{\Gamma, y: \alp}{\alp}$. The cases in part $d)$ where the derivation is either $\rightarrow R$, $\rightarrow L$, or $\Pi^{\Box} L$ must be done by case analysis on whether or not $\tau_1, \tau_2$, or $\msub \tau'$ (respectively) are atomic box-types. This is because the result of $\negD$ and $\posG$ in the induction hypotheses depends on whether or not these assumptions are boxed. If they are boxed they will appear unboxed in $\negD$ and omitted from $\posG$, otherwise they remain in $\posG$. Part $e)$ also involves case analysis on some subgoals, depending on if the assumptions in $\Gamma$ are boxed or not. Some cases in part $f)$ involve multiple case analyses- once on the result of the induction hypothesis, and once on the shape of the assumptions in $\Gamma$. The entire proof is given in \cite{my-thesis}.

\section{Automation tactic}

We present an overview of our automation tactic which implements the focusing calculi from Section \ref{focusingChapter}. We begin with a walk-through of the theorem prover followed by a summary of its performance on various case studies from PL theory.

\subsection{Example}
To demonstrate the capabilities of the (meta-)theorem provers we examine how \bel automatically proves that reduction is closed under expansion, a key lemma needed to show weak-head normalization for the simply-typed lambda-calculus. Specifically, we prove that if term \lstinline[columns=fixed]{M} steps to term \lstinline[columns=fixed]{N} and \lstinline[columns=fixed]{N} reduces to type \lstinline[columns=fixed]{A}, then \lstinline[columns=fixed]{M} does as well.

Building upon the theory presented in Section \ref{theory},  we encode the notion of reducibility using a logical relation by categorizing terms by the type they reduce to. 
We formalize the predicate in \belend's reasoning logic, as it requires a strong, computational function space unlike the weak function space of LF. 
Inductive properties about contextual objects are defined using indexed recursive types \cite{POPL12}. We encode the set of reducible terms using the recursive type \lstinline[columns=fixed]{Reduce}, which is stratified by its index \lstinline[columns=fixed]{tp} \cite{strat-types}. 

{
  \centering
  \begin{minipage}{1\linewidth}
\begin{lstlisting} [xleftmargin=.05\textwidth]
stratified Reduce : {A:[ ⊢ tp]}{M:[ |- term A]} ctype =
| I   : [ ⊢ halts M] -> Reduce [ ⊢ b ] [ ⊢ M]
| Arr : [ |- halts M] 
          -> ({N:[ |- term A]} Reduce [ ⊢ A] [ |- N] -> Reduce [ ⊢ B ] [ |- app M N]) 
            -> Reduce [ ⊢ arr A B ] [ |- M]
;
\end{lstlisting}
 \end{minipage}
}

To begin our theorem, we load the \bel file containing our LF and computation-level signatures into \harend. 
\har then takes as input a theorem name and statement, and index of induction variable specified by its position in the overall goal, counted from left-to-right: 

\begin{lstlisting}[language=Beluga, deletekeywords={of}]
Name of theorem: bwd_closed
Statement of theorem: {A:[ |- tp]} {M: [ ⊢ term A]} {M': [ ⊢ term A]} [ |- step M M'] 
                            -> Reduce [ |- A] [ |- M'] -> Reduce [ |- A] [ |- M]
Induction order (empty for none): 1
\end{lstlisting}

Once loaded, users may simply invoke the tactic
\texttt{auto}\footnote{In \harend, this tactic is invoked using \texttt{inductive-auto-solve}. See \href{https://beluga-lang.readthedocs.io/en/latest/harpoon/proof-automation.html\#inductive-auto-solve}{https://beluga-lang.readthedocs.io/en/latest/harpoon/proof-automation.html\#inductive-auto-solve} for a description on how to use the tactic.} to prove the lemma. Behind the scenes, \bel
gives the specified induction variable to our solver which then
performs induction on said variable and generates the respective
induction hypotheses for each subgoal. It then performs a round of inversions and bounded
depth-first proof search on each produced subgoal. Users have the
option of choosing their own depth bound by providing an argument to
\texttt{auto}, otherwise it is set to 3. The proof search algorithm closely implements the focusing calculus presented in Section \ref{focusingChapter}. As it searches for a proof, the solver also constructs proof terms in the form of LF and computation-level terms, and LF substitutions.  If the solver can prove
each case, the constructed proof term is presented to the user. If it
cannot prove all subgoals, users will have to manually split on the specified variable (with the \texttt{split} tactic) and then call \texttt{auto} on each applicable case. In these instances, since no induction variable is specified, the solver will only perform inversions and bounded search. We show below
the proof script that is being generated by invoking only the \texttt{auto} tactic. 

\begin{center}
  \begin{minipage}{0.9\linewidth}
\begin{lstlisting}[language=Beluga, deletekeywords={of},numbers=left]
proof bwd_closed : {A : ( |- tp)} {M : ( |- term A)}{M' : ( |- term A)} 
                    [ |- step M M'] -> Reduce [ |- A] [ |- M'] -> Reduce [ |- A] [ |- M] =
/ total 1 /
intros
{ A : ( |- tp), M : ( |- term A), M' : ( |- term A)
| x : [ |- step M M'], x1 : Reduce [ |- A] [ |- M']
; solve
    let [ |- Y] = x in
    case [ |- A] of
    | [ |- b] =>
      let (I x2  : Reduce [ |- b] [ |- M']) = x1 in
      let ([ |- halts/m X3 X4] : [ |- halts M']) = x2 in
      let [ |- val/c] = [ |- X4] in I [ |- halts/m (sstep Y X3) val/c]
    | [ |- arr X X1] =>
      let (Arr x2 x3  : Reduce [ |- arr X X1] [ |- M']) = x1 in
      let ([ |- halts/m X5 X6] : [ |- halts M']) = x2 in
      let [ |- val/abs ] = [ |- X6] in
      Arr [ |- halts/m (sstep Y X5) (val/abs )]
        (mlam N => fn y =>
          bwd_closed [ |- X1] [ |- app M N] [ |- app M' N] [ |- stepapp Y] (x3 [ |- N] y)) } ;
\end{lstlisting}    
  \end{minipage}
\end{center}

The \bel program that our solvers construct is given in lines 8 - 20, which gets spliced into the overall proof script. From the given proof script, we can
extract a complete \bel program \cite{harpoon-interactive}. 
Previously, this lemma
could be proved with no less than 10 \har tactic calls
(see also \cite{harpoon-interactive}).

The proof begins with the automatic application of the
\lstinline[columns=fixed]{intros} tactic, which performs the uniform
right phase. The rest of the proof is constructed using \belend's theorem provers. It first unboxes \lstinline[columns=fixed]{x} (line 8), concluding
the uniform left phase. Since an induction variable was specified, the
algorithm immediately splits on the variable. After a split,
there is a round of inversions for each produced subgoal followed by bounded focused proof search. In the first case (\lstinline[columns=fixed]{A = b}) after the inversions (line 13), the computation-level solver focuses on the constructor \lstinline[columns=fixed]{I} and solves its subgoal by providing the appropriate boxed LF proof term (\lstinline[columns=fixed]{ halts/m (sstep Y X3) val/c}) which it finds again through focused proof search, but this time on the LF level. In the second case (\lstinline[columns=fixed]{A = arr X X1}), the solver focuses on \lstinline[columns=fixed]{Arr} which requires solving two subgoals, one of which is of function type. It is here where the solver focuses on the induction hypothesis to solve a subgoal. If the solver focuses on the ``wrong'' assumption, it will execute backtracking and continue focusing until it has exhausted all possible branches on the bounded search tree.

\subsection{Evaluation}
\label{eval}

The proof-search procedures behind our automation tactic is the beginning of the implementation of the focusing calculi presented in Section \ref{focusingChapter}. There are several areas of incompleteness that may be improved in the future \cite{my-thesis}. Nevertheless, the tactic is able to prove a number of interesting theorems both semi- and fully-automatically. We provide a summary of these case studies here.

\begin{table}[h]
    \centering
    \begin{tabular}{m{0.24\linewidth} | m{0.13\linewidth} | m{0.14\linewidth} | m{0.32\linewidth}}
        Case study & Automation & Difficulty & Interesting proof features \\
        \hline \hline
        \text{MiniML/fix type} preservation & Full & Advanced & \text{Solving substitutions,} \text{I.H. appeal, inversions} \\
        \hline
        \text{MiniML/fix value} soundness & Full & Basic & I.H. appeal \\
        \hline
        \text{STLC weak-head} \text{normalization lemmas} & Full & Intermediate & \text{Inversions, I.H. appeal,} \text{higher-order solving}  \\
        \hline
        \text{STLC type} uniqueness & Partial & Basic & I.H. appeal, inversions \\
        \hline
        \text{Untyped $\lambda$-calculus} \text{reduction lemmas} & Full/Partial & Basic & \text{I.H. appeal} \\
    \end{tabular}
    \caption{Overview of case studies.}
    \label{tab:my-label}
\end{table}

We categorize our case studies by the amount of automation that may be successfully applied. We say full automation is applied when only \texttt{auto} is used to complete a proof (as in the previous example). Partial automation is used on induction proofs when not all the subgoals fall within the prover's applicable subset. In these instances, \texttt{auto} is used after a variable split has been manually made, and only on a portion of the subgoals. All proofs proceed by induction along with various features that we have outlined. 

We use our tactic to prove key lemmas required to show weak-head normalization for the simply-typed lambda-calculus. The backwards closed lemma is particularly interesting as it requires higher-order function type solving. Type uniqueness for the simply-typed lambda-calculus can be proven semi-automatically as two of its cases involve parameter variables and context block schemas. We prove all lemmas regarding ordinary reduction for the untyped lambda-calculus automatically, and all but one lemma regarding parallel reduction automatically. These lemmas are used to prove equivalence of the ordinary and parallel reductions, and ultimately the Church-Rosser theorem for each reduction procedure. For MiniML without fixpoints, we are able to prove type preservation and value soundness fully-automatically. Preservation in particular showcases the solvers ability to solve for substitutions \cite{my-thesis}.

\belend's proving power does not yet surpass that of Twelf's. However, \bel is able to reason directly using logical relations, unlike Twelf. Certain properties, like normalization theorems, are most commonly proven using logical relations. In Twelf, users must find alternative proof methods \cite{norm-Twelf, norm-Twelf-Abel}, which may be conceptually different from on-paper formulations and require more work from the user to construct additional machinery. In \belend, such logical relations may be directly translated from on-paper formulations and their proofs become simplified with the use of our automation tactic.

\section{Related work}

\subsection{Proof environments based on HOAS} 
Twelf \cite{twelfSystemDesc} currently provides the most automation
out of all HOAS-based proof assistants designed to formalize
PLs. It fully automatically proves many theorems
such as type-preservation for MiniML, Church-Rosser for the simply
typed lambda-calculus, and cut-admissibility for first-order logic
\cite{Schurmann98cade,Schurmann00phd}. It essentially has a simple
loop that splits on an assumption based on a heuristic,
generates induction hypothesis, and tries to prove a given goal by
bounded depth-first proof search. If the last step fails, it will
again split on an assumption heuristically picking one. This has
proven remarkably effective. However, the kind of properties that can be
stated is more restrictive. For example, Twelf's meta-logic permits
only theorems expressed as $\Pi_2$ statements and lacks support for
recursive types. Therefore it cannot, for example, directly deal with
proofs that proceed via logical relations.  
More importantly, Twelf's prover does not produce proof terms and
therefore provides no way to verify its proofs. Even worse, if the
meta-theorem prover fails to find a proof automatically, there is no
possibility for the user to pick up the pieces and manually proceed. 

Abella is another HOAS-based proof assistant with the capabilities to
mechanize meta-theoretic proofs about PLs
\cite{abellaSystemdesc}. 
Semi-automation exists in Abella
provided in the form of tactics that are similar to the ones
previously implemented in \harend. Contexts and (simultaneous)
substitutions in Abella are not treated as first-class like they are
in \belend. This means that in order to use such constructs, users
must manually define them and any properties they wish to utilize
about them (e.g. context weakening). These can further add to the
complexity of formal theorem proving. 
As Twelf, Abella also does not construct proof objects during proof
development. Moreover, the tactics used in constructing proofs in Abella
concentrate on ``small'' steps and there is presently no analogue to
the proof search tactic that we describe in this paper. However,
we believe that similar semi-automation could be added to Abella to
discharge straightforward cases.

\subsection{General proof environments}

The Hybrid system aims to bring full HOAS support to general proof environments like Coq and Isabelle/HOL \cite{first-hybrid, official-hybrid}. These systems often have more users compared to specialized systems and thus more reason to provide tooling and automation. 
Both systems employ the use of tacticals, which allow users to create their own tactics specific to their mechanizations. 

In Coq, users write decision procedures in a tactic language like LTac \cite{ltac}. There are general automated proof search tactics (auto, eauto, iauto and jauto) but they do not conduct any case analysis (including inversions), inductions, or rewritings, and are intended to finish a proof instead of complete it entirely \cite{coq-book}. Coq also offers many libraries to assist with building and verifying infrastructure needed in PL mechanizations. Autosubst may be used as a library for example to automatically generate parallel substitution operations for a
custom type and prove the related substitution lemmas \cite{autosubst}. The libraries Ott \cite{ott} and LNgen \cite{lngen} may be used together to generate locally nameless
definitions from a specification and provide the corresponding recursion schemes and infrastructure lemmas. However these tools are limited as they only automate trivial lemmas.

For much of its automation, Isabelle elicits the help of several external solvers. Sledgehammer \cite{sh}, for example, takes a goal and heuristically chooses from Isabelle’s libraries containing various lemmas, definitions, and axioms, a few hundred applicable ones to perform search over. Then, translates the goal and each of these assumptions to SMT (first-order logic) and sends the query off to an external SMT or resolution-based solver. In its own system, Isabelle performs various general-purpose proof search methods which help discharge simple parts of a proof allowing users to focus on the main ones \cite{isabelle-auto}. They also have several strengthened endgame tactics which are meant to finish a proof but provide no hints upon failing.

Despite the abundance of tooling in these systems, some specialized
systems (\bel and Twelf for example) offer more automation for PL
mechanizations. This is because these systems have fixed specification
logics, so automatic proof procedures can be more intricate. Hybrid on
the other hand allows for encodings of various specification logics,
therefore proof procedures are more difficult as they must be
customized to work for different logics.

\section{Conclusion}

In closing, we have presented the theorem and meta-theorem provers behind \belend, a specialized proof assistant with sophisticated built-in support for specifying formal systems. These provers perform two-leveled proof search over a core subset of \belend's logic which allows for the automatic completion of many simple lemmas and cases of PL theory proofs. Users of \bel may now bypass simple proofs and focus their energy on the interesting cases. Along with our implementation, we provide a theoretical foundation for our solvers in the form of a cut-free sequent calculus, which is easy to understand, and a sound and complete focusing calculus, which closely reflects our implementation. These provide us with a way to study our implementation and ensure its correctness.


Our next steps are to expand the solver so that its proving capabilities are equivalent to that of the logic presented in this paper \cite{my-thesis}. After that we plan to add support for context block schemas, and substitution and parameter variables, which should bring its proving power up to that of Twelf's. Finally, it would be interesting to extend the focusing calculus (and implementation) with automatic induction, similar to \cite{focus-w-induct}.

\nocite{*}
\bibliographystyle{eptcs}
\bibliography{generic}
\end{document}
